\documentclass[12pt]{article}
\usepackage[top=1in,left=1in,right=1in,bottom=1in]{geometry}

\usepackage{amsmath,amsthm,amssymb,graphicx}
\usepackage{apalike}
\usepackage{breakcites}

\usepackage[margin=1cm,labelfont=bf,
	format=hang,textfont={it}]{caption}

\usepackage[colorlinks=true,urlcolor= blue, 
linkcolor = blue,citecolor=blue]{hyperref}

\usepackage{setspace} 

\newtheorem{theorem}{Theorem}
\newtheorem{remark}{Remark}

\begin{document}

\title{\bf Understanding the population structure correction regression}

\author{
The Tien Mai$^{(1)}$ and Pierre Alquier$^{(2)}$
\\
\small
\\ \small
$^{(1)}$ Oslo Centre for Biostatistics and Epidemiology, Department of Biostatistics, 
\\ \small
University of Oslo, Norway.
\\ \small
$^{(2)}$ RIKEN Center for Advanced Intelligence Project, Tokyo, Japan.
}
\date{email: t.t.mai@medisin.uio.no}

\maketitle

\begin{abstract}
Although genome-wide association studies (GWAS) on complex traits have achieved great successes, the current leading GWAS approaches simply perform to test each genotype-phenotype association separately for each genetic variant. Curiously, the statistical properties for using these approaches is not known when a joint model for the whole genetic variants is considered. Here we advance in GWAS in understanding the statistical properties of the "population structure correction" (PSC) approach, a standard univariate approach in GWAS. We further propose and analyse a correction to the PSC approach, termed as "corrected population correction" (CPC). Together with the theoretical results, numerical simulations show that CPC is always comparable or better than PSC, with a dramatic improvement in some special cases.
\end{abstract}


\section{Introduction}
\label{sec1}

In high dimensional data analysis where the number of covariates $p $ is larger than the number of samples $n$,  penalized regression approaches,  such as Lasso, are one of the most popular approach \cite{hastie2005elements,buhlmann2011statistics,giraud2014introduction,efron2016computer}. However, interpreting the results of the Lasso in terms of hypothesis testing or uncertainty quantification is difficult.

Motivated by genome-wide association studies (GWAS), we focus on the following question: given a response vector $y$ of $n $ samples and a matrix of (genetic) covariates  $X_{n\times p}$ formed by $p $ (genetic) covariates of $n $ samples, we want to determine which covariates associate with the response. Although the variable selection problem is a classical problem in statistics, in this context it is still a big challenge as the number of covariates is huge compared to the sample size, which prohibits the use of the classical methods. Moreover, in many practical situation, the genomic data are huge and can not even be examined on a personal laptop. For example, in human genomics, the number of covariates (SNPs, single-nucleotide polymorphism) as well as the number of samples are often at hundreds thousands  \cite{bycroft2018uk}; or it can be at order tens of millions covariates when using k-mers (an alignment-free biomarker type) as in bacterial genomics \cite{lees2016sequence} with thousands of samples.

Besides the computational reason, most of the theoretical result on the Lasso are on $\ell_2$ estimation of the parameter and not on variable selection. These two objectives are known to be incompatible in general, see~\cite{yang2005can,leeb2008sparse} (and~\cite{zhao2006model} illustrated this in the case of the Lasso). Some results for variable selection were derived for thresholded versions of the Lasso, for example in~\cite{lounici2008sup}, but are valid under strong assumptions that are usually not satisfied in GWAS. Some procedure leading to significance tests and confidence intervals are proposed e.g in \cite{javanmard2014confidence,van2014asymptotically,zhang2014confidence} but they are not easy to handle and costly to compute when using penalized regression such as Lasso.

This lead to the widely use of a univariate model for testing the association of a trait and a covariate (say $X_{\cdot 1} $):
\begin{align*}
y = \alpha_1 X_{\cdot 1} + e
\end{align*}
to estimate $\alpha_1$ and test its significant. However, the omitted variables have an effect, which we will model by multivariate and so the fitted model should be:
\begin{align*}
y = \beta_1 X_{\cdot 1} + \sum_{j>1} \beta_j X_{\cdot j} + \varepsilon.
\end{align*}
The effect of omitted variables (that is, the difference between $\alpha_1$ and $\beta_1$, depends strongly on the dependence between $ X_{\cdot 1} $ and the other covariances. In GWAS data, the covariates (often SNPs) are in some dependent structures which is called \textit{linkage disequilibrium}. This is due to the population structure: many of SNPs have different frequencies in each population \cite{price2006principal,price2010new}.

If one uses a univariate regression and ignores the effect of the other covariates, then they can be effectively modelled as part of the error as $e = \sum_{j>1} \beta_j X_{\cdot j} + \varepsilon $. However the covariates are correlated due to the population structure, this leads to a correlation between the tested covariate, say $X_{\cdot 1} $, and the noise term together with the noise of the samples. These correlations can cause in inflated type-1 error rates \cite{derks2017relation}. To handle this problem, the so-called 'population structure correction' approach had been introduced and successfully applied in practice, see for example \cite{price2006principal,price2010new,lippert2011fast} among others.

In principle, population structure correction is an alternative way to implicitly model the other covariates that are not being tested at the time. This can be done through the latent subspace of these variables. A natural way is to use principal  component analysis to extract some features that contain most information of the other covariates $X_{\cdot -1}$ and use these features as representatives added in the univariate regression of $X_{\cdot 1} $. In this way, it can be seen as a dimension-reduction approach. Another way that is also being the standard approach in GWAS is to use 'linear mixed model' framework in which the covariates that are not directly being tested are treated as random. However, several works had shown that inclusion of $X_{\cdot 1} $ in calculating the principal components can lead to loss in power \cite{listgarten2012improved,yang2014advantages}. This motivates and leads to the popular usage of leave-one-chromosome-out method \cite{lippert2011fast,listgarten2012improved,yang2014advantages}.

Although univariate regression approach with population structure correction has become the state-of-the-art approach in GWAS, there are several numerical works have showed that fitting a penalized multivariate regression can exceed it, e.g \cite{wu2009genome,visscher201710,
buzdugan2016assessing,brzyski2017controlling,lees2020improved}. This can be explained as that using population structure correction can be biased. Moreover, population structure correction very much depends on the added latent features of untested covariates.

In this paper, we study the statistical  properties of the population structure correction when assuming the true underlying model is a multivariate linear regression. More specifically, we derive explicitly the bias and the variance of the population structure correction method. Moreover, we also propose and study a simple version of the leave-one-chromosome-out method, termed as 'corrected population correction'. We show theoretically and empirically that 'corrected population correction' approach reduces the variance compare to the population structure method.

The paper is organized as follow. In Section \ref{scModelMethod}, the model formulation and different methods are presented. The main results on the statistical properties of population structure correction and corrected population correction are given in Section \ref{scStatistcalAnalysis}.  Some numerical simulations are conducted in Section \ref{sc_simulations} and we conclude the paper in the final section.

\section{Model and methods}
\label{scModelMethod}

\subsection{Model}
Given a response vector $y$ of $n$ samples and $p$ covariates $X_{\cdot j}$ with $n \ll p$, we assume that the response vector relates to the covariates by the following linear model
\begin{equation}
 \label{true-model}
 Y_i = \sum_{j=1}^p \beta_j X_{i,j}  + \varepsilon_i, i = 1,\ldots,n
\end{equation}
or, summarized by $Y = X\beta + \varepsilon$ in matrix form with
$$
\left(
\begin{array}{c}
 Y_1 \\
 \vdots \\
 Y_n
\end{array}
\right)
= \left(
\begin{array}{c c c}
 X_{1,1} & \dots  & X_{1,p} \\
 \vdots  & \ddots & \vdots \\
 X_{n,1} & \dots & X_{n,p}
\end{array}
\right)
\left(
\begin{array}{c}
 \beta_1 \\
 \vdots \\
 \beta_p
\end{array}
\right)
+
\left(
\begin{array}{c}
 \varepsilon_1 \\
 \vdots \\
 \varepsilon_n
\end{array}
\right) .
$$
We assume that $ \varepsilon_i \sim \mathcal{N}(0,\sigma^2) $. With a deterministic $X$, this leads to $ \mathbb{E}(Y) = X\beta .  $

The problem is to estimate the coefficient $\beta_j$. Up to a re-ordering of the variables, say that the coefficient is $\beta_1$. For simplicity, there is no intercept: we assume that the variables are already centered, and normalized.

\paragraph{Notations:} For any matrix $A$, $A_{(-j)}$ denotes matrix $A$ without its $j$-th column, and $A_{s:j}$ is the submatrix with only columns $s, s+1, \dots,j$.   We use the same convention for column vectors: $\beta_{s:j}$ means that we extract entries from $s $ to $j $.

\subsection{Population structure correction (PSC)}

The idea is to perform a principal component analysis (PCA) on $X$ and then use some principal components corresponding to the top leading eigenvalues. In other words, 
$$
X^\top X =
\bar{W}^\top
\left(
\begin{array}{c c c}
 \bar{\lambda}_1 & \dots  & 0 \\
 \vdots  & \ddots & \vdots \\
 0 & \dots & \bar{\lambda}_p
\end{array}
\right)
\bar{W}
\text{ where }
\bar{W} = 
\left(
\begin{array}{c c c}
 \bar{W}_{1,1} & \dots  & \bar{W}_{1,p} \\
 \vdots  & \ddots & \vdots \\
 \bar{W}_{p,1} & \dots & \bar{W}_{p,p}
\end{array}
\right)
$$
and $\bar{\lambda}_1\geq \dots \geq \bar{\lambda}_p \geq 0$. The matrix $X^\top X $ is also known as the 'kinship' matrix in genomic research.

Here it will be more convenience to think of PCA as an SVD, that is 
$$ X = \bar{U} \left(
\begin{array}{c c c}
 \bar{\sigma}_1 & \dots  & 0 \\
 \vdots  & \ddots & \vdots \\
 0 & \dots & \bar{\sigma}_r
\end{array}
\right) \bar{V}^\top = \bar{U}\bar{\Sigma} \bar{V}^\top$$
where $\bar{r}={\rm rank}(X)$ and $\bar{\sigma}_1>\dots>\bar{\sigma}_r>0$. It is easy to see that $\bar{V}$ contains the $\bar{r}$ first columns of $\bar{W}$, and so $\bar{V}=\bar{W}$ as soon as ${\rm rank}(\bar{X})=p$.

Then the idea is simply to estimate the model for some $k$,
$$ Y_i = \bar{\alpha} X_{1,i} + \sum_{j=1}^k \bar{\gamma}_j\bar{U}_{i,j}+\bar{e}^{(k)}_i ,$$
or
\begin{align}
\label{PSC.model}
Y = X_{\cdot 1 } \bar{\alpha} + \bar{U}_{(1:k)}\bar{\gamma} + \bar{e}^{(k)},
\end{align}
hoping that $\bar{\alpha}$ is a good proxy for $\beta_1$ in model \eqref{true-model}.

\subsection{Corrected population correction (CPC)}

We propose a modified procedure: first, perform a PCA on $X_{(-1)}$ ($X$ without its first column $X_{\cdot 1}$ is denoted by $X_{(-1)}$), in other words:
$$ 
X_{(-1)} = U \left(
\begin{array}{c c c}
 \sigma_1 & \dots  & 0 \\
 \vdots  & \ddots & \vdots \\
 0 & \dots & \sigma_r
\end{array}
\right) V^\top = U\Sigma V^\top
$$
where $r={\rm rank}(X_{(-1)}) $ and $\sigma_1>\dots>\sigma_r>0$.

Similar to PSC, it is simply to estimate the model, for some $k$,
$$ Y_i = \alpha X_{1,i} + \sum_{j=1}^k \gamma_j U_{i,j} + e^{(k)}_i ,$$
or
\begin{align}
\label{CPC.model}
Y = X_{\cdot 1} \,  \alpha + U_{(1:k)}\gamma + e^{(k)},
\end{align}
and hoping that $\alpha$ is a good proxy for $\beta_1$ in model \eqref{true-model}.

We would like to note that this method is a simple version of the so-called popular method 'leave-one-chromosome-out' in GWAS \cite{yang2014advantages}.

\subsection{Why does PSC need to be corrected?}
In general, the CPC model is not ``correct'' in the sense that $\mathbb{E}(Y)\neq X_{\cdot 1 } \alpha + U_{1:k} \gamma$. In other words, in \eqref{CPC.model} we don't have $\mathbb{E}(e^{(k)})=0$, in general. However, assume that $k={\rm rank}(X_{-1})$, then we have
$$
Y = X_{\cdot 1 } \alpha + U \gamma + e
$$
and note that $X_{(-1)} = U\Sigma V^\top $ leads to $X_{(-1)} V(V^\top V)^{-1}\Sigma^{-1} = U$. Thus, this model is equivalent to~\eqref{true-model} with $\alpha=\beta_1$, by identification:
$$
X_{\cdot 1 } \alpha + X_{(-1)}( V(V^\top V)^{-1}\Sigma^{-1} \gamma ) = X\beta 
= 
X_{\cdot 1 }\beta_1 + X_{(-1)} \left(
\begin{array}{c}
 \beta_2 \\ \vdots \\ \beta_p
\end{array}
\right) 
$$
(and so $\varepsilon = e $ in this case). Therefore, for a well chosen $k$, the model is actually exact.
For this reason, we can reformulate the problem as: with the true model
\begin{equation}
 \label{true-model-2}
Y = X_{\cdot 1 } \alpha + U \gamma + \varepsilon,
\end{equation}
$\gamma\in\mathbb{R}^r$ where $r={\rm rank}(X_{(-1)})$,
what is the effect on $\alpha$ to estimate instead, for some $k$,
\begin{equation}
 \label{wrong-model}
Y = X_{\cdot 1 } \alpha + U_{(1:k)} \gamma_{1:k} + e^{(k)}
\end{equation}
where we actually have $e^{(k)} = U_{(k+1):r} \gamma_{(k+1):r}+\varepsilon$. This is simply a problem of omission of variables: what is the effect of the omission of $U_{(k+1):r}$?

On the other hand, for $k={\rm rank}(X)$, we have
$$ Y = X\beta + \varepsilon = \bar{U} \bar{\Sigma} \bar{V}^\top \beta  + \varepsilon  = \bar{U} (\bar{\Sigma} \bar{V}^\top \beta)  + \varepsilon  = \bar{U} \bar{\gamma} + \varepsilon $$
and so the PSC model
\begin{align}
\label{PC.model}
Y = \bar{\alpha} X_{1} + \bar{U} \bar{\gamma} + \varepsilon
\end{align}
is simply not identifiable (the variable $X_{\cdot 1 } $ is twice in the model). When $k<{\rm rank}(X)$, the model might be identifiable, but the fact that $X_{\cdot 1 } $ is in the first term, and ``partly'' in the second, will lead to a greater bias than in CPC.
This drawback has been figured out in the field of genetic research \cite{yang2014advantages}. For a formal statement see the analysis below.

\section{Statistical Analysis}
\label{scStatistcalAnalysis}

In the following we explicitly derive the bias and the variance for each considered methods above. These results bring insights on understanding how the population structure correction is working practically.  All technical proofs are postponed to Section \ref{sc_proofs}.

\subsection{Main theorems}
We first provide some statistical properties for the CPC method.
\begin{theorem}
\label{theorem:CPC}
Assume that model~\eqref{true-model} or equivalently~\eqref{true-model-2} holds. Then with CPC method we have
	$$  
	{\rm bias}(\hat{\alpha})  
	=
	\frac{ X_{\cdot 1 }^\top U_{(k+1):r} \gamma_{(k+1):r} }{
	X_{\cdot 1 }^\top X_{ \cdot 1 } 
-
\left\|X_{\cdot 1 }^\top U_{1:k} \right\|^2 },
	$$
\begin{align}	
\label{var.of.CPC}
{\rm Var}(\hat{\alpha}) 
	= \frac{\sigma^2}{ X_{\cdot 1 }^\top X_{\cdot 1 } - \left\|X_{\cdot 1 }^\top U_{1:k} \right\|^2} .
\end{align}	
\end{theorem}

It can be seen that $X_{\cdot 1 }^\top  U_{(k+1):r}  \gamma_{(k+1):r} $ measures the correlation between $X_{\cdot 1 }$ and the other part of $X $ (as in the true model~\eqref{true-model-2}) which was not included in the wrong model~\eqref{wrong-model}. Obviously, if the wrong model is actually not ``too wrong'' in the sense that $ \|U_{(k+1):r} \gamma_{(k+1):r} \| \simeq 0 $ then the bias would be small. But when this is not the case, the term is problematic only if $X_{\cdot 1 }$ is correlated with this quantity.

The denominator $ X_{\cdot 1 }^\top X_{ \cdot 1 } 
-
\left\|X_{\cdot 1 }^\top U_{1:k} \right\|^2  $ is just an identifiability term: if $X_{\cdot 1 }$ is {\it too} correlated with the other variables used in model~\eqref{wrong-model},  then the variance of $\hat{\alpha}$ will increase (as usual) but also the bias due to misspecification.

\begin{remark}
As $\gamma$ is unknown in practice, but the $ c_j := X_{\cdot 1 }^\top U_{(j)} $ are observed. So we can give a result under an assumption that depends only on $\gamma$. For example if we assume that $\|\gamma\|_1 \leq B$ (as in the Lasso) then
$$ 
| {\rm bias}(\hat{\alpha}) | 
\leq  
\frac{\|\gamma\|_1 \sum_{j=k+1}^{r} |c_j|}{X_{\cdot 1 }^\top X_{\cdot 1 } -\sum_{j=1}^k c_j^2 } \leq  \frac{B \sum_{j=k+1}^{r} |c_j|}{X_{\cdot 1 }^\top X_{\cdot 1 } -\sum_{j=1}^k c_j^2 } .  
$$
\end{remark}

Statistical properties of the PSC method are given in the following theorem.

\begin{theorem}
\label{thm_PSC}
Assume that model~\eqref{true-model} holds. For the model \eqref{PC.model}, with PSC method, we have
$$
	{\rm bias}(\hat{\bar{\alpha}})  
	=
	\frac{ X_{\cdot 1 }^\top ( \bar{U}_{(k+1):\bar{r}} \bar{\gamma}_{(k+1):\bar{r}} - X_{\cdot 1 } \beta_1) 
	- \beta_1 \left\|X_{\cdot 1 }^\top \bar{U}_{1:k}\right\|^2
	}{  X_{\cdot 1 }^\top X_{\cdot 1 }  - \left\|X_{\cdot 1 }^\top \bar{U}_{1:k}\right\|^2};
$$
\begin{align}
\label{var.of.PC}
{\rm Var}(\hat{\bar{\alpha}}) 
= 
\frac{\sigma^2}{ X_{\cdot 1 }^\top X_{\cdot 1 } - \left\|X_{\cdot 1 }^\top \bar{U}_{1:k} \right\|^2} .
\end{align}
	
\end{theorem}

In the following theorem, we derive the relationship between the variances of these two methods.
\begin{theorem}
\label{thm_compare_variance}
The corrected population correction method reduces the variance of the original population structure correction, i.e
$	
{\rm Var}(\hat{\alpha}) 
	\leq 
{\rm Var}(\hat{\bar{\alpha}}) .
$
\end{theorem}

\begin{remark}
From Theorem \ref{thm_compare_variance},  it states that the corrected population correction (CPC) always returns estimate with smaller variance comparing to the structured population correction (SPC).  From simulations, we conjecture that the biasness of CPC method is also smaller than those from SPC method, however this is not easy to show from our analysis.
\end{remark}

\subsection{Reliable implication check}

Assuming that $ X $ is normalized, that is $ X_{\cdot 1 }^\top X_{\cdot 1 } = 1 $. As we have that
$
| X_{\cdot 1 }^\top U_{(k+1):r} \gamma_{(k+1):r} |
\leq
\sum_{s = k+1}^r | X_{\cdot 1 }^\top U_{s} \gamma_{s} |,
$
and using Cauchy-Schwarz inequality yields $| X_{\cdot 1 }^\top U_{s} |^2 \leq \| X_{\cdot 1 } \|^2 \|  U_{s} \|^2 \leq 1 $ and we obtain $| X_{\cdot 1 }^\top U_{(k+1):r} \gamma_{(k+1):r} |
\leq
\sum_{s = k+1}^r | \gamma_{s} |. $
Thus, we have
$$
\left| {\rm bias}(\hat{\alpha})  \right|
	\leq
\frac{ \left\| \gamma_{(k+1):r}  \right\|_1  }{  \left|  1 - \left\|X_{\cdot 1 }^\top U_{1:k} \right\|^2 \right| } 
:= \frac{N}{D} , 
\quad
	{\rm Var}(\hat{\alpha}) \leq \frac{\sigma^2}{ \left| 1 - \left\|X_{\cdot 1 }^\top U_{1:k} \right\|^2 \right| } : = \frac{\sigma^2 }{D}.
$$
Now, CLT yields $ \hat{\alpha} \sim \mathcal{N} ( \mathbb{E}\left(\hat{\alpha}) , {\rm Var} (\hat{\alpha}) \right) $. We want to test the null hypothesis that
$$
H_0: \alpha = 0.
$$
Under this null hypothesis, we have 
$ 
| \mathbb{E}(\hat{\alpha})  | \leq \frac{N}{D}
$ 
and $(1-a)100\% $ confidence interval is
$
\mathbb{E}(\hat{\alpha}) 
\in
\left[ \hat{\alpha} - z_{\frac{a}{2}} \sqrt{{\rm Var}(\hat{\alpha})}  ,   
\hat{\alpha} + z_{\frac{a}{2}} \sqrt{ {\rm Var}(\hat{\alpha})}    \right]
$
or
$
\hat{\alpha}
\in
 \left[ - \mathbb{E}(\hat{\alpha})  - z_{\frac{a}{2}} \sqrt{{\rm Var} (\hat{\alpha})} 
  , 
 \mathbb{E}(\hat{\alpha})  + z_{\frac{a}{2}} \sqrt{ {\rm Var}(\hat{\alpha})}    \right]
$. Thus we obtain 
\begin{align*}
\hat{\alpha}
\in
 \left[ - \frac{N}{D}  - z_{\frac{a}{2}} \sqrt{\frac{\sigma^2}{D} } 
  , 
  \frac{N}{D}  + z_{\frac{a}{2}} \sqrt{\frac{\sigma^2}{D} }    \right],
\end{align*}
where $z_{\frac{a}{2}} = \Phi^{-1}(1- \frac{a}{2}) $ and $\Phi^{-1}(\cdot)$ is the normal cumulative  distribution function. 

The above analysis lead to the following tests for the null hypothesis: 
\begin{align*}
{\rm If}\quad \hat{\alpha}
\notin
 \left[ - \frac{N}{D} - z_{\frac{a}{2}}\sqrt{\frac{\sigma^2}{D} } 
  , 
  \frac{N}{D} + z_{\frac{a}{2}} \sqrt{\frac{\sigma^2}{D} }    \right]\quad {\rm  then \,\, reject} \,\, H_0.
\end{align*}

If the noise variance $ \sigma $ is not known, one can use a consistent estimate $\hat{\sigma} $, for example as in \cite{dicker2014variance,sun2012scaled,
janson2017eigenprism} and the confidence interval would become 
\begin{align*}
\hat{\alpha}
\in
 \left[ - \frac{N}{D}  - t_{(n-1,\frac{a}{2})} \sqrt{\frac{\hat{\sigma}^2}{D} } 
  , 
  \frac{N}{D}  + t_{(n-1,\frac{a}{2})} \sqrt{\frac{\hat{\sigma}^2}{D} }    \right],
\end{align*}
where the normal cumulative distribution function is replaced by the student distribution.

\section{Numerical simulations}
\label{sc_simulations}
\subsection*{Setup}

In this section, we investigate basic properties of the PSC and CPC methods studied above. We fix $p=100, n = 1000 $ for low dimension setting and $p=1000, n = 600 $ for high dimension setting. The noise variance is fixed at $\sigma^2 = 1 $. 

We generate the parameter $\beta \in \mathbb{R}^p $ such that its first component $\beta_1 $ is fixed to 1, and other non-zero components was sampled uniformly at random from $\{ \pm 1\} $. The sparsity of $\beta $ will be changed in each setting corresponding to $\|\beta \|_0 = 20,  100 $. The response $Y$ is simulated as in linear model \eqref{true-model}. 

For each setting, we simulated 100 independent datasets and report the average results together with their standard deviations.  The number of principal components $ k $ added in models \eqref{CPC.model} and  \eqref{PSC.model} are varied from 1 to 30.

\subsection*{Example: worst case scenario for PSC}
Here we show cases that PSC does not work well while CPC performs superior results. 

\begin{figure*}[!h]
\caption{Summary the structured X: When $X_{\cdot -1}$ is removed, a principal component is also removed.}
\label{fgSTRCsummary}
\includegraphics[scale=.5]{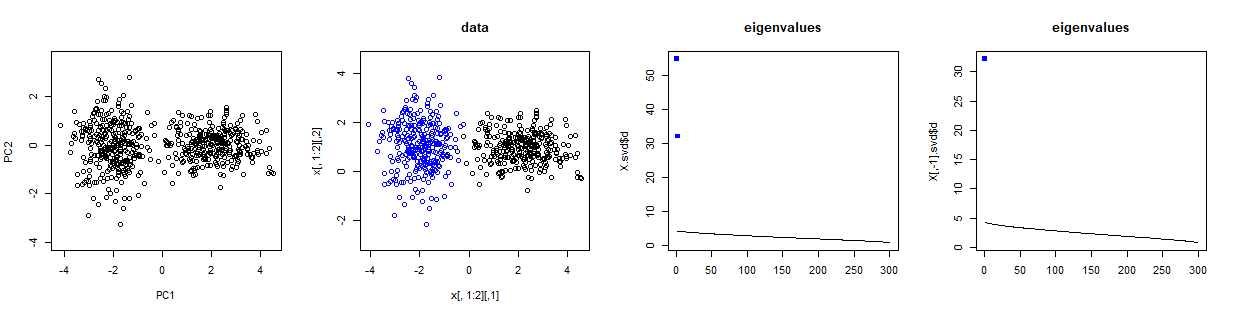}
\end{figure*}

We consider the structured $X$ such that its first two columns $X_{\cdot 1}$ and $X_{\cdot 2}$ are corresponding to its first two leading principal components. A brief summary of the data can be found in the Figure \ref{fgSTRCsummary}.

\begin{figure}[!ht]
\caption{Structured $X$. Estimates of $\beta_1$ with different number of Principal components (PCs)}
\label{fgstructure}
\centering
\includegraphics[scale=0.45]{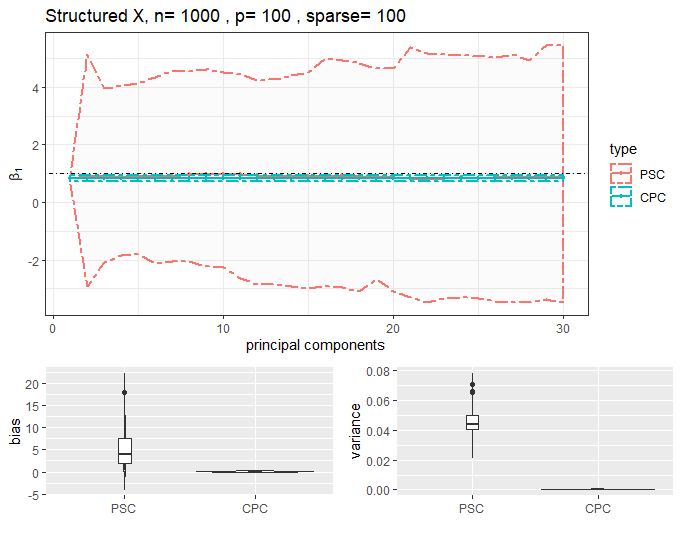}
\includegraphics[scale=0.45]{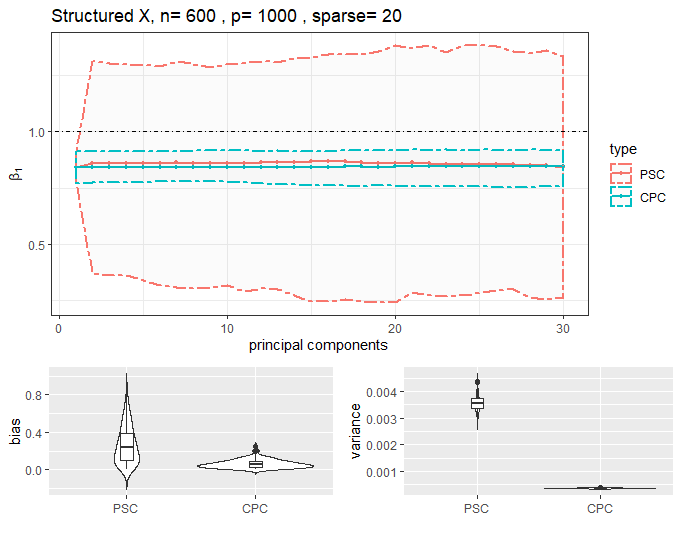}
\end{figure}

In this case, it is clear to see that PSC can actually be very biased whereas CPC is very stable and accurate, see Figure \ref{fgstructure}. This example demonstrates that including $X_1$ (the covariate being tested) in the calculation of the principal components can be very harmful.

\subsection*{Behaviour in other cases}
We further consider the following settings for the design matrix:
\begin{itemize}
\item Independent $X$: In this setting, $X_{ij} \sim \mathcal{N} (0,1) $.
\item Dependent $X$: We consider $X_{i\cdot} \sim \mathcal{N} (0, \Sigma) $, where $ \Sigma_{ij} = 0.5^{|i-j|} $.
\item Binary $X$: the $X_{ij}$ is simulated from the set $\{\pm 1\} $ with equal probability.
\end{itemize}

\begin{figure*}[!ht]
\caption{Independent $X$. Estimates of $\beta_1$ with different number of Principal components (PCs)}
\label{fgindependent}
\centering
\includegraphics[scale=0.45]{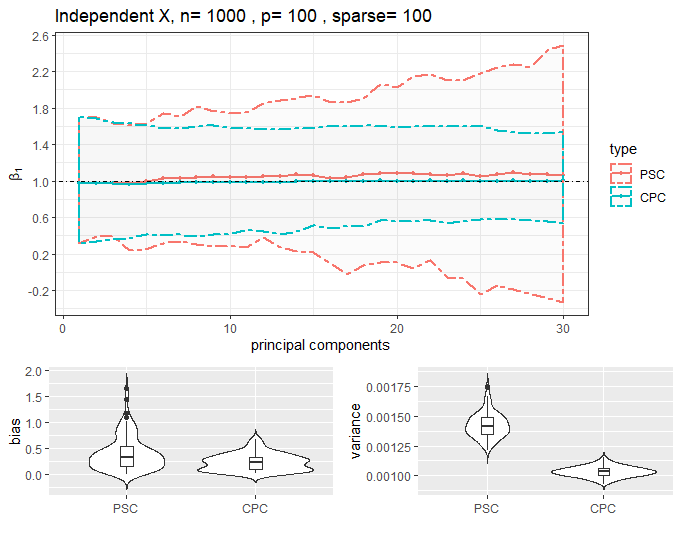}
\includegraphics[scale=0.45]{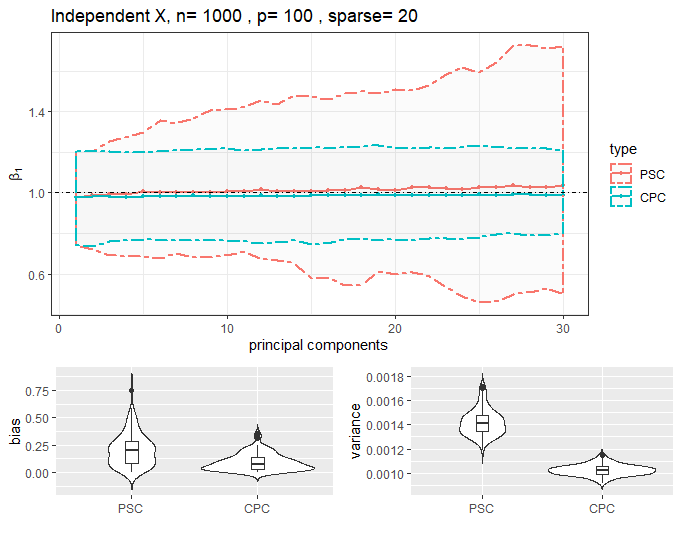}

\centering
\caption{Dependent $X$. Estimates of $\beta_1$ with different number of Principal components (PCs)}
\includegraphics[scale=0.45]{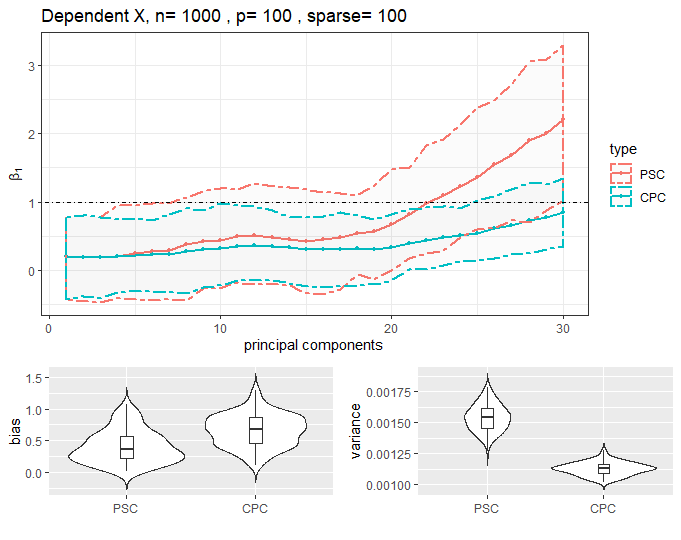}
\includegraphics[scale=0.45]{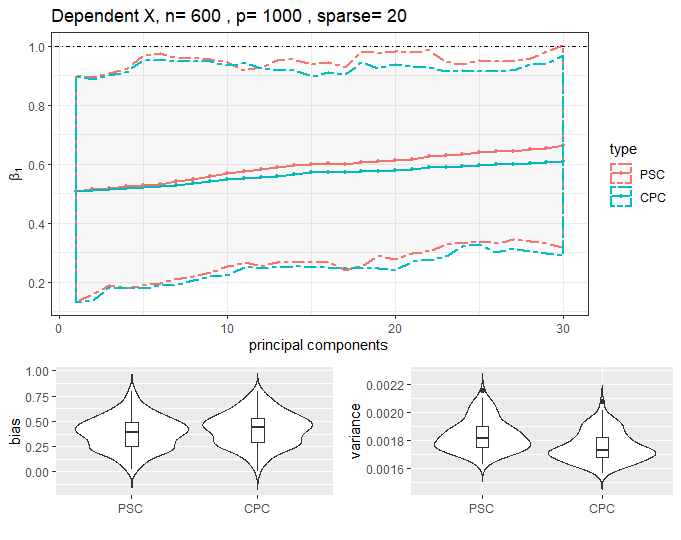}
\label{fgdepent}

\caption{Binary $X$. Estimates of $\beta_1$ with different number of Principal components (PCs)}
\label{fgBina}
\centering
\includegraphics[scale=0.45]{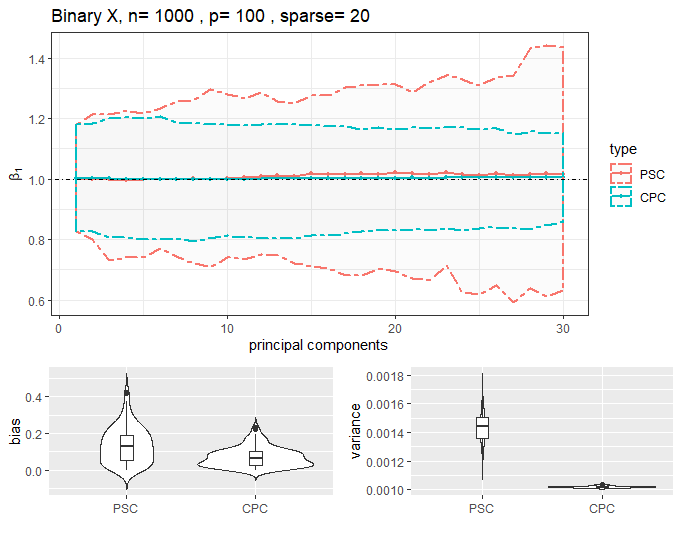}
\includegraphics[scale=0.45]{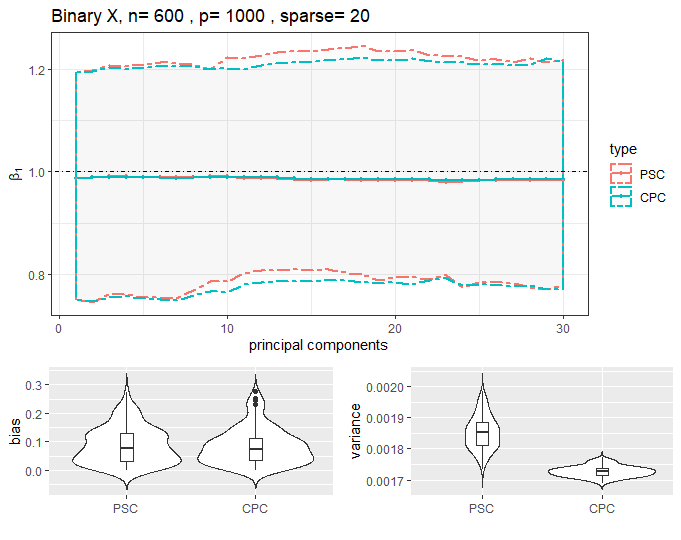}
\end{figure*}

Results from simulations, Figures \ref{fgindependent},\ref{fgdepent} and \ref{fgBina} confirm our theoretical results above. In general, the CPC method performs similarly to PSC method. However, CPC return the results with less variation than the PSC method. Moreover, the PSC method is very much depending on $ k $, the number of principal components added in the model.

\subsection*{Real data assessment in a wheat GWAS data}

We apply two methods to a real wheat GWAS data which is available in the R package 'BGLR' \cite{perez2014genome}. The data consists of 599 wheat lines: lines (responses) were evaluated for grain yield and each line has been genotyped with 1279 markers.

\begin{figure*}[ht]
\caption{Histogram of the absolute errors and relative errors of 1279 covariates in wheat data with 10 principal components.}
\label{realdata_histo}
\centering
\includegraphics[width=15cm, height=6cm]{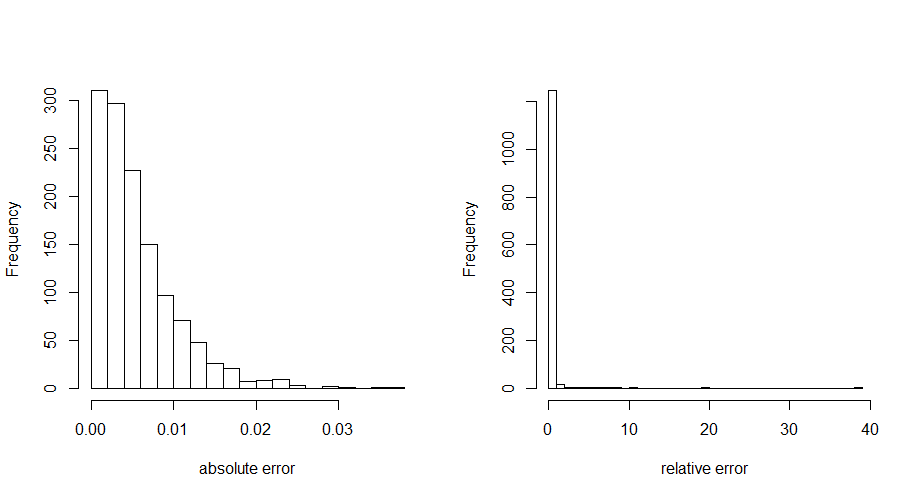}
\end{figure*}

We run CPC and PSC across 1279 covariates and report the absolute errors
$$
| \hat{\beta}_j^{CPS} -\hat{\beta}_j^{PSC} |
$$
and the relative errors
$$
\frac{ | \hat{\beta}_j^{CPC} -\hat{\beta}_j^{PSC} | }{ |\hat{\beta}_j^{PSC} |}.
$$
These results are given in Figure \ref{realdata_histo}. 

Regarding the histogram in Figure \ref{realdata_histo}, the conclusion is clear: for most coefficients, PSC and CPC lead to similar estimation, but for some of them, the deviation is extremely high. There are in total 55 covariates such that their relative errors are greater than 0.5 (and there are 33 covariates such that their relative errors are greater than 1).  Therefore,   including the covariate being tested in the calculation of the principal components could create a huge difference.

\section{Closing Discussion}
In this paper, we have dicussed the statistical properties of the widely used method,  structure population correction method, in genome-wide association studies.  We have also proposed and studied a simple version of the  'leave-one-chromosome-out' in GWAS, termed as Corrected population correction method.  Our theoretical analysis and simulations show that the structure population correction method (although efficient computationally) should be used with more careful as it comes with higher variance due to model-mispecification.  The corrected population correction method,  which requires higher computational cost,  returns better results as it avoids model-mispecification.

\section*{Acknowledgments}
T.T.M would like to thank Jukka Corander and John A Lees for useful discussion on GWAS. The research of T.T.M was supported by the European Research Council (SCARABEE project) no. 742158.

\section*{Conflict of interest}
The authors declare no potential conflict of interests.

\section*{Availability of data and materials}
The R codes and data used in the numerical experiments are available at:  
\\
\url{https://github.com/tienmt/understand_SPC} .

\clearpage
\appendix
\section{Proofs}
\label{sc_proofs}

\subsection{Proof of Theorem \ref{theorem:CPC} for CPC}
\subsubsection*{Analysis of the bias}
In~\eqref{wrong-model}, let's define the matrix $ Z = \left(X_{\cdot 1 } | U_{1:k} \right) $, put $\zeta = (\alpha|\gamma_{i:k}^\top)^\top$, then it becomes:
$$
Y = Z \zeta + e^{(k)}
$$
and the least square estimator is given by
\begin{align}
\label{least_square_estimator}
\hat{\zeta} = (Z^\top Z)^{-1} Z^\top Y 
=
 (Z^\top Z)^{-1} Z^\top [Z \zeta + e^{(k)}] 
=
 \zeta + (Z^\top Z)^{-1} Z^\top e^{(k)}.
\end{align}
So the bias of this estimator is simply
$$
{\rm bias}(\hat{\zeta}) = \mathbb{E} \left( \hat{\zeta} \right) - \zeta = (Z^\top Z)^{-1} Z^\top \mathbb{E}(e^{(k)}),
$$
that is
\begin{align}
\label{bias.CPC}
{\rm bias}(\hat{\zeta}) = (Z^\top Z)^{-1} Z^\top U_{(k+1):r} \gamma_{(k+1):r}.
\end{align}
Let us now make this more explicit. First,
$$
 Z^\top U_{(k+1):r} \gamma_{(k+1):r}
 =
 \left(
 \begin{array}{c}
  X_{\cdot 1 }^\top U_{(k+1):r} \gamma_{(k+1):r}
  \\ 0 \\ \vdots \\ 0
 \end{array}
 \right),
$$
and
$$
Z^{T} Z = \left(
\begin{array}{c c c c}
X_{\cdot 1 }^\top X_{\cdot 1 }  & X_{\cdot 1 }^\top U_{(1)} & \dots & X_{\cdot 1 }^\top U_{(k)} \\
X_{\cdot 1 }^\top U_{(1)}  & 1 & & \\
\vdots & & \ddots & \vdots \\
X_{\cdot 1 }^\top U_{(k)} & 0 & \dots & 1
\end{array}
\right).
$$
Now we know that the bias of $\hat{\zeta}$ in \eqref{bias.CPC} satisfies
$$
(Z^{T} Z ) {\rm bias}(\hat{\zeta})  =  Z^\top U_{(k+1):r} \gamma_{(k+1):r} ,
$$
and it can be written explicitly as
\begin{align*}
\left(
\begin{array}{c c c c}
X_{\cdot 1 }^\top X_{\cdot 1 }  & X_{\cdot 1 }^\top U_{(1)} & \dots & X_{\cdot 1 }^\top U_{(k)} 
\\
X_{\cdot 1 }^\top U_{(1)}  & 1 & \dots   &  0 
\\
\vdots & & \ddots & \vdots 
\\
X_{\cdot 1 }^\top U_{(k)} & 0 & \dots & 1
\end{array}
\right)
\left(
\begin{array}{c}
{\rm bias}(\hat{\alpha})
\\ {\rm bias}(\hat{\gamma}_1) \\ \vdots \\ {\rm bias}(\hat{\gamma}_k)
\end{array}
\right)
=  \left(
\begin{array}{c}
X_{\cdot 1 }^\top U_{(k+1):r} \gamma_{(k+1):r}
\\ 0 \\ \vdots \\ 0
\end{array}
\right).
\end{align*}
The generic equation in the second part of the system is
$$ X_{\cdot 1 }^\top U_{(j)} {\rm bias}(\hat{\alpha}) + {\rm bias}(\hat{\gamma}_j) = 0 $$
yielding
$$ {\rm bias}(\hat{\gamma}_j) = - X_{\cdot 1 }^\top U_{(j)} {\rm bias}(\hat{\alpha}) .$$
Plugging this into the first equation:
\begin{align*}
X_{\cdot 1 }^\top X_{\cdot 1 } {\rm bias} (\hat{\alpha}) +  \sum_{j=1}^k X_{\cdot 1 }^\top U_{(j)}    {\rm bias}(\hat{\gamma}_j) 
=  
X_{\cdot 1 }^\top U_{(k+1):r} \gamma_{(k+1):r}
\end{align*}
gives
$$
\left[ X_{\cdot 1 }^\top X_{\cdot 1 }  - \sum_{j=1}^k \left(X_{\cdot 1 }^\top U_{(j)} \right)^2  \right] {\rm bias}(\hat{\alpha})
=  
X_{\cdot 1 }^\top U_{(k+1):r} \gamma_{(k+1):r}.
$$
Thus, we obtain
\begin{align*}
{\rm bias}(\hat{\alpha})
=
 \frac{ X_{\cdot 1 }^\top U_{(k+1):r} \gamma_{(k+1):r}}{ X_{\cdot 1 }^\top X_{\cdot 1 }  - \sum_{j=1}^k \left(X_{\cdot 1 }^\top U_{(j)} \right)^2}
 =
  \frac{ X_{\cdot 1 }^\top U_{(k+1):r} \gamma_{(k+1):r}}{X_{\cdot 1 }^\top X_{\cdot 1 } 
	-
	 \left\|X_{\cdot 1 }^\top U_{1:k} \right\|^2}.
\end{align*}

\subsubsection*{Variance analysis of CPC}

From \eqref{least_square_estimator}, assuming that $ {\rm Var}(\epsilon) = \sigma^2 $, we have 
\begin{align*}
{\rm Var}(\hat{\zeta}) 
=  
(Z^\top Z)^{-1} Z^\top {\rm Var}(u^{k}) Z  (Z^\top Z)^{-1}
= 
 \sigma^2 (Z^\top Z)^{-1},
\end{align*} 
or
\begin{align*}
Z^\top Z {\rm Var}(\hat{\zeta}) 
=    \sigma^2 \mathbf{I}.
\end{align*} 
As we are only interested in estimating the variance of $ \hat{\alpha} $, from the above formula we obtain
\begin{align*}
\begin{cases}
(X_{\cdot 1 }^\top X_{\cdot 1 })  {\rm Var}(\hat{\alpha}) + \sum_{j=1}^{k} Cov(\hat{\alpha}, \hat{\gamma_j})  X_{\cdot 1 }^\top U_{(j)} = \sigma^2,
\\
(X_{\cdot 1 }^\top U_{(j)})  {\rm Var}(\hat{\alpha}) + Cov(\hat{\alpha}, \hat{\gamma_j}) = 0.
\end{cases}
\end{align*}
Substituting the second equation into the first one to get
\begin{align*}
{\rm Var}(\hat{\alpha}) = \frac{\sigma^2}{ X_{\cdot 1 }^\top X_{\cdot 1 } - \sum_{j=1}^{k} ( X_{\cdot 1 }^\top U_{(j)})^2 }.
\end{align*}

\subsection{Proof of Theorem \ref{thm_PSC} for PSC}
\subsubsection*{Analysis of the bias}
In \eqref{PSC.model}, let's define the matrix $ \bar{Z} = \left(X_{\cdot 1 } | \bar{U}_{1:k} \right) $, put $\bar{\zeta} = (\bar{\alpha}|\bar{\gamma}_{i:k}^\top)^\top$, and use the least square estimator:
\begin{align}
\label{PC.least.sq}
\hat{\bar{\zeta}} 
= 
(\bar{Z}^\top \bar{Z})^{-1} \bar{Z}^\top Y
\end{align}
and all we have is that
$$
Y = \bar{U} \bar{\gamma} + \varepsilon.
$$
Thus, we have
$$
 \mathbb{E}(\hat{\bar{\zeta}}) = \mathbb{E}\left[ (\bar{Z}^\top \bar{Z})^{-1} \bar{Z}^\top Y \right] = (\bar{Z}^\top \bar{Z})^{-1} \bar{Z}^\top  \bar{U} \bar{\gamma}
$$
or, 
\begin{equation}
 \label{step-pc}
(\bar{Z}^\top \bar{Z}) \mathbb{E}(\bar{\zeta})
= \bar{Z}^\top  \bar{U} \bar{\gamma}.
\end{equation}
First,
$$
\bar{Z}^{T} \bar{Z} 
= \left(
\begin{array}{c c c c}
X_{\cdot 1 }^\top X_{\cdot 1 } & X_{\cdot 1 }^\top \bar{U}_{(1)} & \dots & X_{\cdot 1 }^\top \bar{U}_{(k)} \\
X_{\cdot 1 }^\top \bar{U}_{(1)}  & 1 & & \\
\vdots & & \ddots & \vdots \\
X_{\cdot 1 }^\top\bar{U}_{(k)} & 0 & \dots & 1
\end{array}
\right)
$$
and then
$$
\bar{Z}^{T} \bar{U}
= \left(
\begin{array}{c c c c c c}
 & & X_{\cdot 1 }^\top\bar{U} & & &  \\ \hline
 1 & \dots & 0 & 0 & \dots & 0 \\
 \vdots & \ddots & \vdots & \vdots & \ddots & \vdots \\
 0 & \dots & 1 & 0 & \dots & 0
 \end{array}
\right)
\Rightarrow
\bar{Z}^{T} \bar{U} \bar{\gamma}
 = \left(
 \begin{array}{c}
  X_{\cdot 1 }^\top\bar{U} \bar{\gamma}
  \\
  \bar{\gamma}_1
  \\
  \vdots
  \\
  \bar{\gamma}_k
 \end{array}
 \right).
$$
Equation~\eqref{step-pc} above becomes:
$$
\left\{
\begin{array}{l l}
X_{\cdot 1 }^\top X_{\cdot 1 }  \mathbb{E}(\hat{\bar{\alpha}}) + \sum_{i=1}^k X_{\cdot 1 }^\top\bar{U}_{(i)} \mathbb{E}(\hat{\bar{\gamma}}_i) & =   X_{\cdot 1 }^\top\bar{U} \bar{\gamma}
\\
X_{\cdot 1 }^\top\bar{U}_{(1)} \mathbb{E}(\hat{\bar{\alpha}}) + \mathbb{E}(\hat{\bar{\gamma}}_1) & = \bar{\gamma}_1
\\
\quad \quad \vdots & \vdots
\\
X_{\cdot 1 }^\top\bar{U}_{(k)} \mathbb{E}(\hat{\bar{\alpha}}) + \mathbb{E}(\hat{\bar{\gamma}}_k) & = \bar{\gamma}_k.
\end{array}
\right.
$$
The generic equations, for $1\leq j\leq k$, can be rewritten as
$$
X_{\cdot 1 }^\top\bar{U}_{(j)} \mathbb{E}(\hat{\bar{\alpha}}) + \mathbb{E}(\hat{\bar{\gamma}}_j)  = \bar{\gamma}_j ,
$$
giving
$$
\mathbb{E}(\hat{\bar{\gamma}}_j) =  \bar{\gamma}_j - X_{\cdot 1 }^\top\bar{U}_{(j)} \mathbb{E}(\hat{\bar{\alpha}}).
$$
Plugging this into the first equation gives
$$
\mathbb{E}(\hat{\bar{\alpha}}) 
\left[ X_{\cdot 1 }^\top X_{\cdot 1 }  - \sum_{i=1}^k \left( X_{\cdot 1 }^\top\bar{U}_{(i)} \right)^2 \right] 
+ 
 \sum_{i=1}^k  X_{\cdot 1 }^\top\bar{U}_{(i)}\bar{\gamma}_i =  
X_{\cdot 1 }^\top\bar{U} \bar{\gamma},
$$
that is
$$
\mathbb{E}(\hat{\bar{\alpha}})
 \left[ X_{\cdot 1 }^\top X_{\cdot 1 } - \sum_{i=1}^k \left( X_{\cdot 1 }^\top\bar{U}_{(i)} \right)^2 \right] 
 + 
  X_{\cdot 1 }^\top \bar{U}_{1:k} \bar{\gamma}_{1:k} 
 =  
 X_{\cdot 1 }^\top \bar{U} \bar{\gamma}
$$
and thus
$$
\mathbb{E}(\hat{\bar{\alpha}}) 
\left[ X_{\cdot 1 }^\top X_{\cdot 1 } - \sum_{i=1}^k \left( X_{\cdot 1 }^\top \bar{U}_{(i)} \right)^2 \right] 
= 
 X_{\cdot 1 }^\top \bar{U}_{(k+1):\bar{r}} \bar{\gamma}_{(k+1):\bar{r}}.
$$
Finally,
$$
\mathbb{E}(\hat{\bar{\alpha}})  
= 
\frac{ X_{\cdot 1 }^\top \bar{U}_{(k+1):\bar{r}} \bar{\gamma}_{(k+1):\bar{r}} }{  X_{\cdot 1 }^\top X_{\cdot 1 }  - \left\|X_{\cdot 1 }^\top \bar{U}_{1:k}\right\|^2} .
$$
Now for the bias,
\begin{align*}
{\rm bias} (\hat{\bar{\alpha}})
& = \mathbb{E}(\hat{\bar{\alpha}})-\beta_1
\\
& = \frac{ X_{\cdot 1 }^\top ( \bar{U}_{(k+1):\bar{r}} \bar{\gamma}_{(k+1):\bar{r}} - X_{\cdot 1 } \beta_1) 
	+
	 \beta_1 \left\|X_{\cdot 1 }^\top \bar{U}_{1:k}\right\|^2
}{  X_{\cdot 1 }^\top X_{\cdot 1 }  - \left\|X_{\cdot 1 }^\top \bar{U}_{1:k}\right\|^2}.
\end{align*}

\subsubsection*{Variance of PSC}
From \eqref{PC.least.sq}, we have
\begin{align*}
{\rm Var}(\hat{\bar{\zeta}} )
 = 
(\bar{Z}^\top \bar{Z})^{-1} \bar{Z}^\top {\rm Var}(Y) \bar{Z} (\bar{Z}^\top \bar{Z})^{-1} 
 = 
(\bar{Z}^\top \bar{Z})^{-1}  \sigma^2
\end{align*}
or
\begin{align*}
(\bar{Z}^\top \bar{Z}) {\rm Var}(\hat{\bar{\zeta}} )
= 
\sigma^2 \mathbf{I}.
\end{align*}
As we are only interested in estimating the variance of $ \hat{\bar{\alpha}} $, from the above formula we obtain
\begin{align*}
\begin{cases}
(X_{\cdot 1 }^\top X_{\cdot 1 }) {\rm Var}(\hat{\bar{\alpha}}) + \sum_{j=1}^{k} Cov(\hat{\bar{\alpha}}, \hat{\bar{\gamma}}_j)  X_{\cdot 1 }^\top U_{(j)} = \sigma^2,
\\
(X_{\cdot 1 }^\top U_{(j)}) {\rm Var}(\hat{\bar{\alpha}}) + Cov(\hat{\bar{\alpha}}, \hat{\bar{\gamma}}_j) = 0.
\end{cases}
\end{align*}
Substituting the second equation into the first to get
\begin{align*}
{\rm Var}(\hat{\bar{\alpha}}) = \frac{\sigma^2}{ X_{\cdot 1 }^\top X_{\cdot 1 } - \sum_{j=1}^{k} ( X_{\cdot 1 }^\top \bar{U}_{(j)})^2 }.
\end{align*}

\subsection{Proof for Theorem \ref{thm_compare_variance}}
\begin{proof}[Proof for Theorem \ref{thm_compare_variance}.]	
We have that
\begin{align*}
	\left\| X_{(-1)}^\top u \right\|^2
	=
	\left\| ( \mathbf{0} , X_{2},\ldots,X_{p})^\top u \right\|^2
	\leq
	\left\| X^\top u \right\|^2 .
\end{align*}
And from the definition that
	$ \bar{U_j} = \arg \max_{\|u\|^2 = 1 } \left\| X^\top u \right\|^2  $, we get 
\begin{align*}
	\left\| X^\top U_{1:k} \right\|^2 
	\leq
	\left\| X^\top \bar{U}_{1:k} \right\|^2 
\end{align*}
	or
\begin{align*}
\left\| X_{\cdot 1 }^\top U_{1:k} \right\|^2  
+
\sum_{j=2}^{p} \left\| X_{(j)}^\top U_{1:k}\right\|^2 
	&	
	\leq
	\left\| X_{\cdot 1 }^\top \bar{U}_{1:k} \right\|^2  
	+ \sum_{j=2}^{p}\left\| X_j^\top \bar{U}_{1:k} \right\|^2 ,
\\
	\left\| X_{\cdot 1 }^\top U_{1:k} \right\|^2  + \max_{\|x\|_2^2 = 1} \sum_{j=2}^{p} \left\| X_{(j)}^\top x_{1:k} \right\|^2 
	&	\leq
	\left\| X_{\cdot 1 }^\top \bar{U}_{1:k} \right\|^2  
	+ \sum_{j=2}^{p}\left\| X_j^\top \bar{U}_{1:k} \right\|^2 .
\end{align*}
	By taking $ x = \bar{U}_{1:k} $, we obtain
\begin{align*}
	\left\| X_{\cdot 1 }^\top U_{1:k} \right\|^2  +  \sum_{j=2}^{p} \left\| X_{(j)}^\top \bar{U}_{1:k} \right\|^2 
	\leq
	\left\| X_{\cdot 1 }^\top \bar{U}_{1:k} \right\|^2  
	+ 
\sum_{j=2}^{p}\left\| X_j^\top \bar{U}_{1:k} \right\|^2. 
\end{align*}
	And thus
	\begin{align*}
	\left\| X_{\cdot 1 }^\top U_{1:k} \right\|^2  
	\leq
	\left\| X_{\cdot 1 }^\top \bar{U}_{1:k} \right\|^2  .
	\end{align*}	
	This yields the conclusion of the theorem.
\end{proof}

\section{Simulation results in other cases}

\begin{figure*}[!ht]
\caption{Binary $X$. Estimates of $\beta_1$ with different number of Principal components (PCs)}
\centering
\includegraphics[scale=0.45]{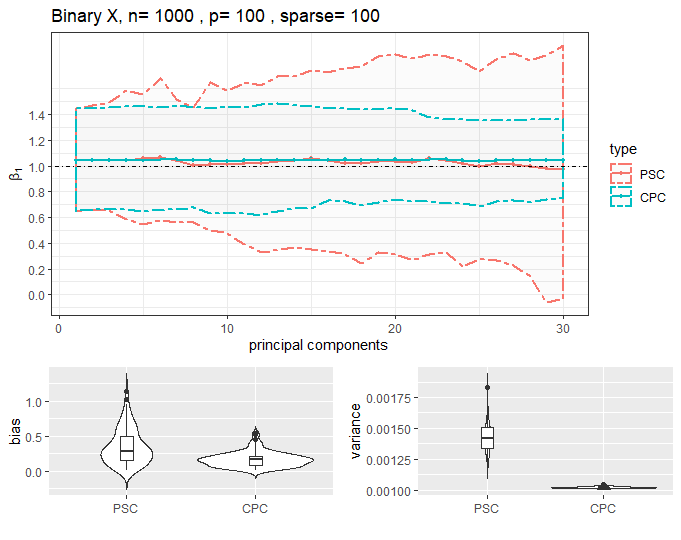}
\includegraphics[scale=0.45]{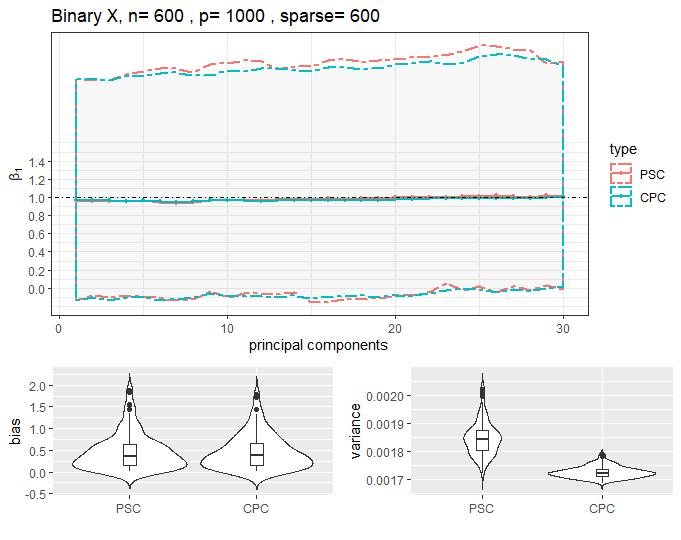}
\end{figure*}

\begin{figure*}[!ht]
\caption{Structured $X$. Estimates of $\beta_1$ with different number of Principal components (PCs)}
\centering
\includegraphics[scale=0.45]{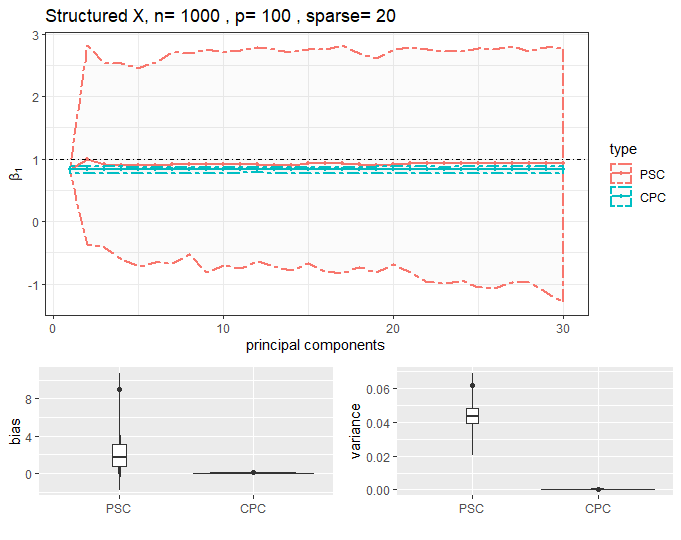}
\includegraphics[scale=0.45]{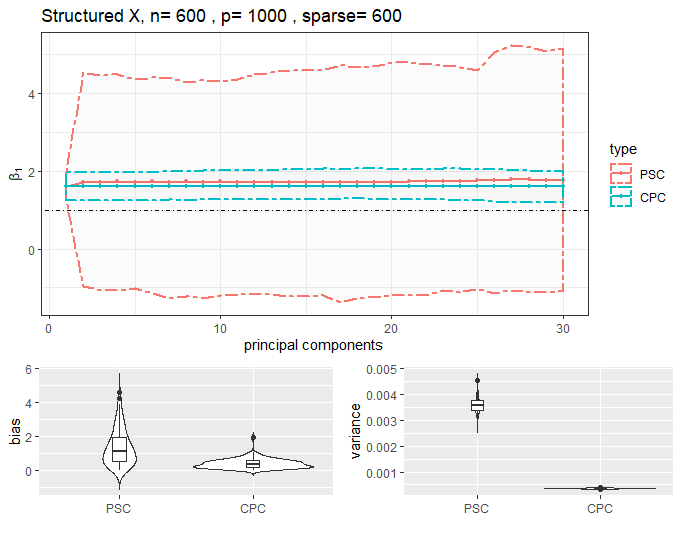}

\caption{Independent $X$. Estimates of $\beta_1$ with different number of Principal components (PCs)}
\centering
\includegraphics[scale=0.45]{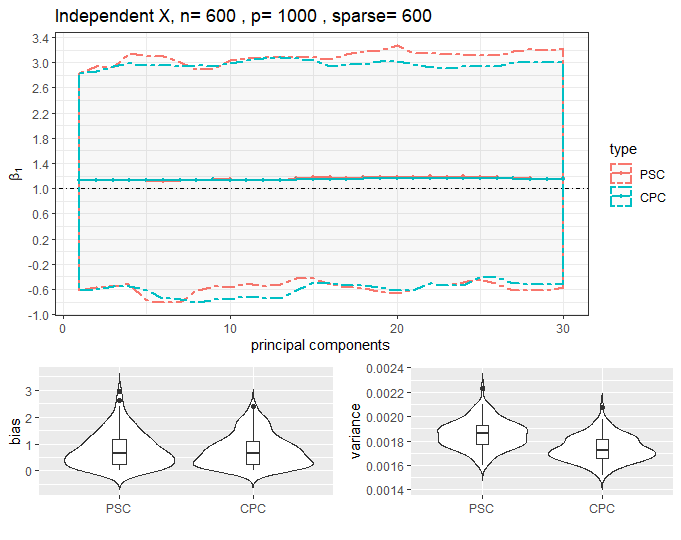}
\includegraphics[scale=0.45]{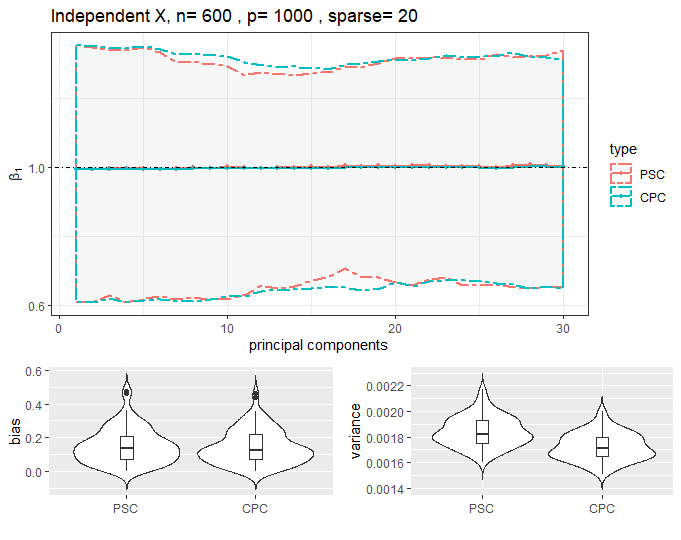}
\caption{Dependent $X$. Estimates of $\beta_1$ with different number of Principal components (PCs)}
\centering
\includegraphics[scale=0.45]{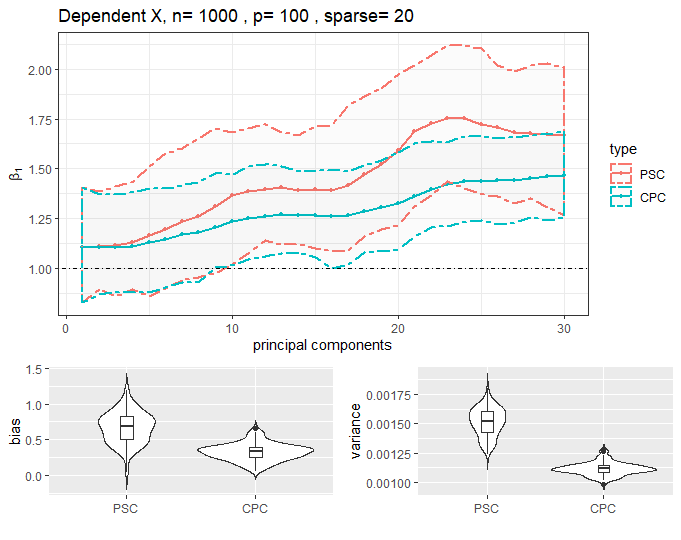}
\includegraphics[scale=0.45]{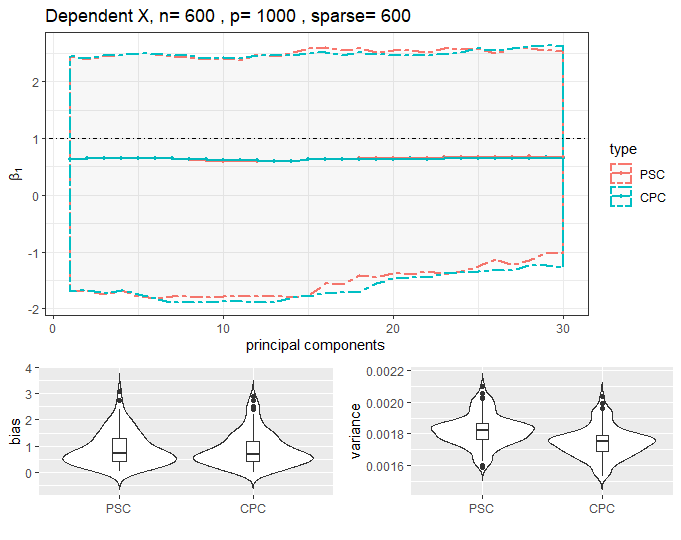}
\end{figure*}

\clearpage
\bibliographystyle{apalike}

\end{document}